\documentclass[12pt]{article}
\usepackage{graphicx}

\newif\ificonv
\iconvfalse   

\def\be{\begin{equation}}
\def\ee{\end{equation}}
\def\bea{\begin{eqnarray}}
\def\eea{\end{eqnarray}}

\newsavebox{\uuunit}
\sbox{\uuunit}
    {\setlength{\unitlength}{0.825em}
     \begin{picture}(0.6,0.7)
        \thinlines
        \put(0,0){\line(1,0){0.5}}
        \put(0.15,0){\line(0,1){0.7}}
        \put(0.35,0){\line(0,1){0.8}}
       \multiput(0.3,0.8)(-0.04,-0.02){12}{\rule{0.5pt}{0.5pt}}
     \end {picture}}

\newif\ifpdf
\ifx\pdfoutput\undefined
   \pdffalse
 \else
   \pdfoutput=1
   \pdftrue
  \usepackage[pdftex]{hyperref}
\fi

\begin{document}
\begin{titlepage}

\font\cmss=cmss10 \font\cmsss=cmss10 at 7pt \leftline{\tt hep-th/0511045}

\vskip -0.5cm \rightline{\small{\tt KUL-TF-05/25}}
\rightline{\small{\tt UB-ECM-PF-05/27}}

\vskip .7 cm

\hfill \vspace{18pt}
\begin{center}
{\Large \textbf{String  splitting  \\ and strong coupling meson decay}}
\end{center}

\vspace{6pt}
\begin{center}
{\large\textsl{A. L. Cotrone $^{a}$, L. Martucci $^{b}$ and W. Troost $^{b}$}}

\vspace{25pt}
\textit{(a) Departament ECM, Facultat de F{\'\i}sica, Universitat de Barcelona and\\ Institut de
Fisica d'Altes Energies,
Diagonal 647, E-08028 Barcelona, Spain}\\
\textit{(b) Institute for Theoretical Physics, K.U. Leuven,\\ Celestijnenlaan 200D, B-3001 Leuven,
Belgium}\\
\vspace{4pt}

\end{center}

\vspace{12pt}

\begin{center}
\textbf{Abstract}
\end{center}

\vspace{4pt} {\small \noindent We study
the decay of high spin mesons using the gauge/string theory correspondence.
The rate of the process is calculated by studying the splitting of a macroscopic string intersecting a D-brane.
The result is applied to the decay of mesons in ${\cal N}=4$ SYM with a small number of flavors and in a gravity dual of large $N$ QCD.
In QCD the decay of high spin mesons is found to be heavily suppressed in the regime of validity of the supergravity description.

}

\vspace{5cm}

\vfill \vskip 5.mm \hrule width 5.cm \vskip 2.mm {\small \noindent e-mail: cotrone@ecm.ub.es, luca.martucci,walter.troost@fys.kuleuven.be}

\end{titlepage}


\section{Introduction}

Since string theory is hoped to describe all fundamental interactions in a unified framework,
it is imperative to try and connect it as much as possible to the well known phenomena of, in particular, particle physics.
The standard description of strongly interacting particles is via constituent quarks that interact according to QCD.
The interaction among constituents is treated perturbatively. This method has met considerable success,
culminating in the description of the physics of jets in high energy collisions of hadrons as well as in the hadron production
in lepton annihilation processes. An important phenomenological tool in matching the basic quark processes (Feynmann diagrams)
to data is the concept of `fragmentation functions', parameterizing how quarks become hadrons. The fact that these fragmentation
functions remain unreachable for an ab initio computation in QCD is indicative of our level of understanding.

In recent years interest has been growing to investigate what string theory could teach us about this realm.
String theory  comes with a bag full of tools, among which the many dualities offer hope that strongly interacting problems
may become accessible by dualising them to weakly coupled ones. In addition, models are being constructed with the string
theory tools that, while not necessarily being entirely realistic, certainly do share a growing number of properties with
the world of quarks as we know it.
The exploration of scattering processes both deep inelastic and hard \cite{PolchinskiStrassler},
has been followed by  studies of quark spectroscopy (see for example \cite{myers1}).

In this paper we address ourselves to another aspect of hadron physics, viz. the decay of high spin mesons into mesons.
Certainly, this is a strong coupling problem in QCD which goes even further than the computation of fragmentation functions.
While one can not expect this question to be amenable to a general realistic treatment at the present time, there are
nevertheless limiting cases where string theory models allow an approximate treatment.
In this way, even if the quark model construction and/or the particular kinematical limits under consideration are certainly not entirely realistic, one might
hope to gain some experience and develop relevant techniques to tackle more realistic cases later.

In the context of the gauge/string correspondence it is possible to study glueball and meson states with large spin.
In this letter we study the decay of mesons, leaving the glueball decay to a brief discussion.
Given a background dual to a gauge theory, mesonic degrees of freedom are usually introduced by placing a small number $N_f$ of ``flavor D-branes'' on which the quarks are living.
If the number of flavors is taken to be much smaller than the number $N$ of colors, the backreaction of these branes on the geometry is ignored: they are considered just as probes \cite{katz1}.
This corresponds to the quenched approximation of QCD.
The regime where the geometry can be trusted is then the usual one, namely large $\lambda=g_s N$
(the 't Hooft coupling) with $N$ very large and $g_s$ very small, but with the additional condition that $N_f<<N$.

In this setting the mesons are seen as the excitations of the flavor D-branes.
We are concerned with high spin excitations, where a meson is semi-classically described as an open string with end-points on these flavor branes.
The mass of the quarks is usually related, in a model-dependent way, to the position of the branes in a ``radial'' direction of the background.
Branes at different such positions represents the possibility of having quarks of different masses in the theory.
The string representing a meson with both quarks of the same mass $m_Q$ is a string with both end-points on the same brane (which we will denote as the $Q$-brane). This string can at some point intersect another brane ($q$-brane) related to quarks of smaller mass $m_q$.
Thus, the string can split in two, each with end-points on the two different branes, as depicted in figure 1.
This is the decay process we are going to study\footnote{For the decay of light mesons, see for example \cite{ss}.}.

\begin{figure}
\begin{center}
\scalebox{0.7}{\includegraphics{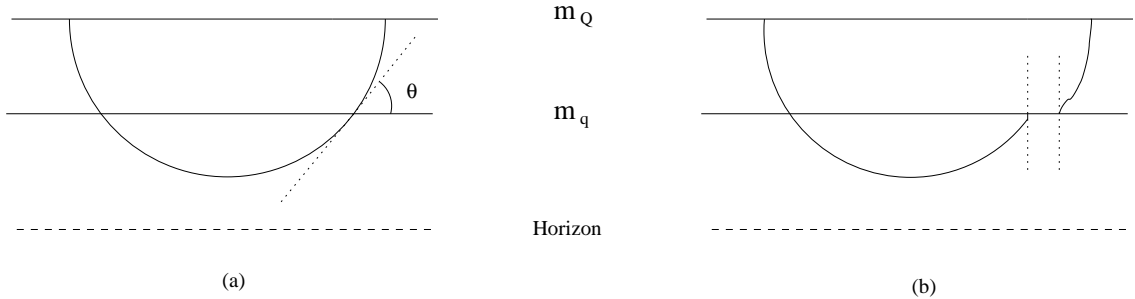}}
\caption{(a) A large spin meson, bound state of two quarks of large mass $m_Q$, described by a string with both end-points on the $Q$-brane. The string intersects the $q$-brane, corresponding to lighter quark masses $m_q$, at an angle $\theta$. In (b), the strings after the splitting, representing two meson bound states of a heavy quark and a light quark.}
\label{figura1}
\end{center}
\end{figure}

From the field theory point of view the main process (amplitude of order $g_s$ with $1\gg g_s \gg 1/N$) is
\begin{equation}\label{decay}
\bar Q Q \rightarrow \bar Q q + Q\bar q \ ,
\end{equation}
where the initial meson is characterized by the spin $S_{\bar Q Q}$ and has an energy $E_{\bar Q Q}=E_{\bar Q Q}[S_{\bar Q Q},m_Q]$.
At the semiclassical level the final state energies ($E$), spins ($S$) and orbital momenta are  in principle
{\em completely determined} by $S_{\bar Q Q}$, $m_Q$ and $m_q$, with no phase space for the decay products\footnote{Note that we are neglecting the contribution to the
spin/energy coming from the back-reacted background fields, which is always of higher order in $g_s$.}.

Two of the main features of the process are that the decay of mesons is {\em highly constrained}, since the splitting can be at only one point, and {\em may not happen}
if the lighter quark mass is too small, as it is evident from figure 1.
At strong coupling the energy of a meson made by a heavy quark and a light quark {\em increases} as the mass $m_q$ of the light quark {\em decreases}.
As a consequence, for the decay to remain possible, the spin of the heavy meson must {\em increase}.
We will discuss this phenomenon more in the following sections\footnote{These and other types of strong coupling phenomena have been studied for example in \cite{katz2,myers1,strassler}.}.

Since the decay process (the splitting of the string) is essentially local and the curvature of the geometry is very small, one can estimate the decay rate by studying the process in flat space.
Thus, in section \ref{sec2} we derive the decay rate of a string intersecting at a generic angle a Dp-brane in flat space.
This result is applied in section \ref{sec3} to study the decay of high spin mesons in the simplest model, namely the ${\cal N}=2$ theory obtained by adding quarks to ${\cal N}=4$ SYM \cite{katz1}.
We examine how the result varies with the parameters, and conclude with a discussion in section \ref{sec4}, which considers also the behavior of the decay rate in confining models, focusing on the example of the large $N$ dual of QCD of \cite{myers2}.
In the same context, we comment on the decay rate of glueballs with high spin.\\


\section{The splitting rate}\label{sec2}

If we consider high spin mesons, a decay following (\ref{decay}) can be described
in a purely geometrical way from the holographically dual point of view. The highly excited meson is represented by
a semiclassical spinning open string with end-points attached to a given brane. The position of this brane fixes the quark mass.
The string can split if it intersects (see figure \ref{figura1}) with another brane that corresponds to less massive quarks.
The intersection will in general be at an angle $\theta\neq \pi/2$.
When the string splits, the two new strings end perpendicularly to the $q$-brane, locally eliminating the transversal tension.

A precise computation of the decay rate seems to be out of reach, due to the non-triviality of the  supergravity
backgrounds we are interested in.
But, since the  process is essentially local, and the curvature of the background is small, one may give a quantitative estimate of the decay rate by considering a simplified
setting in flat space  where an extended string snaps due to the presence of a D-brane.
As an extra bonus, the results we will obtain in this way can be useful in other
contexts where the string splitting due to the presence of branes can be relevant, as for example
in cosmological models that include cosmic F-strings.

In the case of a general Dp-brane with $p>1$, the setting can be obtained by slightly modifying the model discussed in \cite{jjp}.
In order to use the S-matrix formalism, we consider a flat spacetime of the form ${\emph R}\times T^2_\theta\times T^{p-1}_{||}\times T^{8-p}_{\perp}$,
where ${\emph R}$ refers to the time direction.
The  $T^2_\theta$ is parameterized by
$(x^1,x^2)\simeq (x^1+n_1l_1+n_2l_2\cos\theta,x^2+n_2 l_2\sin\theta)$, with $n_1,n_2\in Z$, and contains
the nontrivial geometrical information concerning the angle of incidence.
It is the compactification of the plane where the interaction takes place (the plane in figure 1).
The Dp-brane is then wrapping $T^{p-1}_{||}$ and filling the direction $x^1$ inside $T^2_\theta$,
and these directions will be decompactified at the end of the calculation.
Note that in the actual models of decaying mesons, the $x^2$ direction will be in the transverse geometry and not in the Minkowski part of the metric, which will be accounted for by $x^1$ and by some dimensions of the (decompactification of the) $T^{p-1}_{||}$ factor.

Let us now introduce the macroscopic string,
winding it along
the other periodic direction on $T^2_\theta$, namely $x^2$. The trick of \cite{jjp}, that we adopt here,
is to choose the periodic direction as a ``temperature'' direction,
giving the opposite GSO projection of the usual one, for which the ground states are scalars. Since we are considering very long and therefore very massive strings, one expects that the difference with respect to the usual GSO projection is irrelevant, because it involves a finite number of excitations only.
With this simplification, following \cite{dp,DaiPolchinski}, one can compute the decay rate of the closed string due to the splitting
on the brane from the following correlator of two closed string vertex operators on the disk
\begin{equation}\label{corr}
{\cal A}=\langle {\cal V}_{(0,0)}(p_L,p_R){\cal V}_{-1,-1}(p_L^\prime,p_R^\prime)\rangle\ .
\end{equation}
Even if this is an open string process on the disk, the states are closed strings since they wrap a compactified dimension.
Moreover, the correlator above allows, by means of an optical theorem, to sum over all the possible final states of the splitting,
giving the {\emph{total}} decay rate as the properly normalized imaginary part of the `forward' amplitude.
The ground state vertex operators in the $(-1,-1)$ and $(0,0)$ pictures are given by \cite{jjp}
\bea
{\cal V}_{(-1,-1)}&=&\frac{\kappa}{2\pi\sqrt{V}}:e^{-\phi-\tilde\phi+ip_L\cdot X+ip_R\cdot \tilde X}:\ ,\cr
{\cal V}_{(0,0)}&=&\frac{\kappa}{2\pi\sqrt{V}}\frac{\alpha^\prime}{2}(\psi\cdot p_L)(\tilde\psi\cdot p_R)
:e^{ip_L\cdot X+ip_R\cdot \tilde X}:\ ,
\eea
where the volume factor is $V=\sin\theta l_1 l_2 V_{\perp}V_{||}$, with $V_{||}={\rm Vol}(T^{p-1}_{||})$
and $V_\perp={\rm Vol}(T^{8-p}_{\perp})$. The left and right momenta have to satisfy
 $p_L^2=p_R^2=\frac{2}{\alpha^\prime}$, $p_{L,R}=p\pm L/2\pi\alpha^\prime$, $L=(0,l_2\cos\theta,l_2\sin\theta,0,\ldots)$.
Since at the end we will take the decompactification limit for the directions wrapped by the Dp-brane\footnote{The space orthogonal to both the string and the D-brane, with volume $V_{\perp}$, will be kept compact.},
we also allow for a possible (approximately continuous) velocity of the string along  $T_{||}^{p-1}$. This is the key difference
with respect to the case considered in \cite{jjp}: the string does not move in a direction orthogonal to the brane but in directions
parallel to the brane and orthogonal to the string world-sheet itself\footnote{For example, if we think of the branes in figure 1 as extending in the direction perpendicular to the plane of the screen/paper, the string is moving in that direction.}. The string momentum has the form
\bea
p=\frac{m}{\sqrt{1-v^2}}(1,0,0,\vec v,\vec 0)\ ,
\eea
where $m^2=(l_2/2\pi\alpha^\prime)^2-2/\alpha^\prime$ is the mass of the state, $\vec v \in T_{||}^{p-1}$ and  $\vec 0\in T_{\perp}^{8-p}$.
At leading order in $g_s$, the Dp-brane is  a fixed background object and does not recoil, so in (\ref{corr}) one can take $|\vec v|=|\vec v^\prime|$.

The  amplitude can be obtained by a calculation similar to that for D1-branes in \cite{jjp}.
There is a difference in the Mandelstam variables, due to the different number of Dirichlet conditions in the open string metric.
The relevant limit is that of large $l_2$, hence (unless $\theta=\pi/2$)
large $\sigma\simeq \alpha^\prime(l_2/2\pi\alpha^\prime)^2\cos^2\theta$, and zero value for
$t\equiv -4/\alpha^\prime -2p_L\cdot p_L^\prime$. Actually, we keep $t\neq 0$ as a regulator for the divergence
in the real part of the amplitude as $t\rightarrow 0$ (see also \cite{DaiPolchinski}).
The conformal field theory computation along the lines of \cite{jjp} leads to the following
integral expression, which
is only slightly different from the one for D1-branes:
\be
\int_0^1 dx\, (1-x)^{-1-\alpha't/2}(1+x)^{1+2\sigma-\alpha't/2} x^{-1-\sigma}\ \sim \  2^{2\sigma} \frac{\Gamma(-\alpha't/4) \Gamma(-\sigma)}{\Gamma(-\alpha't/4-\sigma)}\ ,
\ee
where the approximate expression is valid as $t\rightarrow 0$.

The large $\sigma$ limit fluctuates wildly on the real axis, as one can see from the approximate expression,
since it contains the closed string state poles at integer values of $\sigma$ with zero width.
These fluctuations are averaged by taking the limit in a direction in the complex $\sigma$ plane at a small angle.
The infinitely narrow poles will then contribute with the proper weight to the imaginary part.
Practically, this amounts to applying Stirling's formula with the proper choice of phase, and results in
\bea\label{ampl}
{\cal M}\simeq -N_{D^2}\frac{\kappa^2}{(2\pi)^2 V}\frac{4(\sigma)^{1+\alpha^\prime t/4}}{\alpha^\prime t}e^{-i\pi t\alpha^\prime/4}\ ,
\eea
where  the normalization $N_{D^2}$ can be obtained by T-duality from the
standard partition function normalization $2\pi^2 V_9\tau_9$, giving $N_{D^2}=2\pi^2 l_1 V_{||}\tau_p=
2\pi^2 l_1 V_{||}/(2\pi)^{p} (\alpha^\prime)^{(p+1)/2}g_s$.
Taking the imaginary part, the singularity at $t=0$ is resolved, and we obtain
\bea
{\rm Im} {\cal M}=\frac{g_s l_2}{32\pi^2}\times\frac{\cos^2\theta}{\sin\theta \alpha'^{3/2}}
\times
\frac{(2\pi\sqrt{\alpha^\prime})^{8-p}}
{V_\perp}\ .
\eea

Since we are considering the case of macroscopic strings with $l_2\gg \sqrt{\alpha^\prime}$ and mass $m\simeq l_2/2\pi\alpha^\prime$,
the  decay rate can be written in the following final form
\bea\label{rate}
\Gamma=\frac{1}{m}{\rm Im}{\cal M}= \frac{g_s}{16\pi\sqrt{\alpha^\prime}}\cdot\frac{\cos^2\theta}{\sin\theta}
\left(\frac{V_\perp^{(min)}}{V_\perp} \right)\ ,
\eea
where $V_\perp^{(min)}=(2\pi\sqrt{\alpha^\prime})^{8-p}$.
Note that this quantity is finite and does not depend on the (transversal) velocity of the string, since it is computed in the rest frame of the latter.

The interpretation of the decay rate (\ref{rate}) is the following. If we think in string units, the decay rate is
the product of three contributions. First of all, we have the natural $V_\perp^{(min)}/{V_\perp}$
suppression given by the transversal torus.
It just states that the distance between the string and the brane in the transverse directions should be of order $\alpha'$ in order for the interaction to take place.
Second, we have the factor $1/\sin\theta$ that is completely
consistent with the idea that a string intersecting a brane can snap, and as it becomes more and more parallel to the brane the breaking probability increases, since the tension of the string creates a bigger transversal force which helps the string splitting.
Third, we have the $\cos^2\theta$ term which is the natural term symmetric as $\theta\rightarrow -\theta$ that vanishes for the supersymmetric configuration  $\theta=\pi/2$, for which the string does not split\footnote{Note that the calculation above is not valid, strictly speaking, for the extreme values $\theta=0$ (by construction) and $\theta=\pi/2$ (when we are no more in the Regge regime of large $\sigma$). In the latter case the behavior of the resulting rate (\ref{rate}) is nevertheless the expected one.}.

\section{Holographic meson decay}\label{sec3}

Let us now use the general results obtained above to study some dynamical process in the context of the gauge/string duality.
The kind of process we are interested in concerns gauge theories with flavors in the fundamental representation of the gauge group. As shown in \cite{katz1}, when the number of flavors is small (in contrast to the number of colors which is taken very large in the 't Hooft limit), we can study them from the holographic point of view by adding appropriate probe branes in the supergravity background dual to the theory without flavors.

Let us try to be somewhat general. The simplest supergravity backgrounds dual to 4d field theories have metrics of the form\footnote{The generalization to other dimensions and to the thermalized case is trivial.}
\bea\label{metric}
ds^2=e^{A(r)}(-dt^2+d\rho^2+\rho^2d\eta^2+dx_3^2)+e^{B(r)}dr^2+G_{ij}(r,\phi)d\phi^id\phi^j\ ,
\eea
where $r$ and $\phi^i$, $i=1,\ldots,5$, describe respectively the radial and the angular coordinates of  the six dimensional internal space, and $r$ is associated (in a model dependent way) to the energy scale of the dual theory whose UV and IR regimes correspond to large and small $r$ respectively. Quite generally, we can think that the probe D-brane associated to the addition of a flavor $Q$ partly fills some internal angular directions $\chi^a$ while it is located at a fixed value of the remaining ones $\psi^I_Q$. Furthermore, it fills the radial coordinate from $r=\infty$ up to a point defined by  fixed angles $\chi^a_Q$ and a  minimal radius $r_Q$. This minimal radius is then  holographically associated to the flavor mass $m_Q$ in a model dependent way. The fluctuations of the brane describe low mass mesonic states, while mesonic states with very large spin can be described by semiclassical spinning open strings with end-points attached to the flavor D-brane.

We will now focus on the high spin mesons associated to rigid spinning strings whose world-sheet is of the form
\bea
t=\tau\ ,\ \eta=\omega\tau\ ,\ r=r(\sigma)\ ,\ \rho=\rho(\sigma)\ ,\ \chi^a=\chi^a_Q\ ,\ \psi^I=\psi^I_Q\ .
\eea
The relevant equations of motion can be easily derived from the effective action
\bea\label{effact}
S=-\frac{1}{2\pi\alpha^\prime}\int d\tau d\sigma e^A(r)\sqrt{(1-\omega^2\rho^2)[(\rho^\prime)^2+e^{B(r)-A(r)}(r^\prime)^2]}\ ,
\eea
supplemented by the boundary conditions $r|_{\partial\Sigma}=r_Q$ and $\rho^\prime|_{\partial\Sigma}=0$ \cite{myers1}.
For our purposes, it is convenient to  fix the remaining reparameterization invariance by choosing the gauge $\sigma=r$, so that the only effective dynamical field is $\rho(\sigma)$. Then, if $r_0$ indicates the minimal radius reached by the string, the energy and the spin of the meson are given by $E=2F[r_0,r_Q]$ and $J=2H[r_0,r_Q]$ where
\bea
F[a,b]&=&\frac{1}{2\pi\alpha^\prime}\int_{a}^b d\sigma e^{A(\sigma)}\sqrt{\frac{(\rho^\prime)^2+
e^{B(\sigma)-A(\sigma)}}{1-\omega^2\rho^2}}\ ,\cr H[a,b]&=&\frac{\omega}{2\pi\alpha^\prime}\int_{a}^b d\sigma \rho^2
e^{A(\sigma)}\sqrt{\frac{(\rho^\prime)^2+ e^{B(\sigma)-A(\sigma)}}{1-\omega^2\rho^2}}\ .
\eea
One can  in principle invert these relations in order to get for example $E$ and $r_0$ as functions of the spin (and $r_Q$).

Let us consider now the effect of the introduction of another D-brane associated to a lighter flavor $q$ of mass  $m_q<m_Q$, i.e. with $r_q<r_Q$. We want to study the possible decay of the above string, associated to a meson schematically of the kind $\bar Q Q$, into a couple of strings representing the mesons $\bar Q q$ and $\bar q Q$. We will focus on decay rates that can be described within the semiclassical picture, where the string classically intersects the $q$-brane and then can split in the semiclassical regime. For other kinds of mesons, whose D-brane is not aligned with the brane corresponding to the heavier meson in such a way that the spinning string intersects it, the meson decay involves world-sheet instantonic transitions and then is exponentially suppressed in the semiclassical regime.
In order not to have this exponential suppression we must then fulfill the conditions that $r_0[J]\leq r_q$ and $\psi^I_q=\psi^I_Q$.
If these conditions are satisfied,  our classical string can split into two open strings with end-points attached to different branes which indeed correspond to mesons of the kind ${\bar Q}q$ and ${\bar q}Q$. Rigid spinning strings of this kind were studied in \cite{pt} and in our case we expect our states to be some excited version of these rigidly rotating strings, with also some linear momentum.

Even if we will not determine the explicit form of the outcoming strings, it is important to note that their energies and total angular momenta (computed with respect to the rest frame of the initial meson) are completely determined by the classical picture\footnote{See \cite{iengo1,plefka} for examples involving closed strings.}.  Indeed we can immediately conclude that the lightest outcoming meson will have energy $E_1=F[r_q,r_Q]$ and total angular momentum $J_1=H[r_q,r_Q]$, while the heavier meson will have energy $E_2=E-E_1$ and angular momentum $J_2=J-J_1$.
The outcoming states will also have definite and opposite linear momenta. If for example  $P^1$ and  $P^2$ denote the linear momenta of the lightest outcoming meson in the directions\footnote{These are the coordinates in the metric (\ref{metric}), not to be confused with the $x^{1,2}$ of the previous section, that here correspond to $\rho$ and $r$ respectively.} $x_1=\rho\sin\omega\tau$ and $x_2=\rho\cos\omega\tau$, we have that
\bea
P^1(t)&=& \frac{\omega}{2\pi\alpha^\prime}\int_{r_q}^{r_Q} d\sigma \rho\cos\omega\tau e^{A(\sigma)}\sqrt{\frac{(\rho^\prime)^2+
e^{B(\sigma)-A(\sigma)}}{1-\omega^2\rho^2}}\ ,
\eea
and $P^2$ can be obtained by the same expression by replacing  $\cos$ with $-\sin$.

Let us now see what we can say in general on the rate for such a decay using the results obtained previously.
Two basic ingredients we need are the velocity $v$ of the string in the point where it splits and its angle $\theta$ with the brane. These are given by
\bea
v=\omega\rho(r_q)\quad,\quad \cos^2\theta=\frac{(\rho^\prime(r_q))^2}{e^{B(r_q)-A(r_q)}+(\rho^\prime(r_q))^2}\ .
\eea

In order to determine the decay rate, we have also to take into account the  suppression due to the effective transverse volume. This is a delicate point since 
such a transversal quantum delocalization of the string can be infinite \cite{susskind}.
In fact, if the string is free to sit at a generic point of a transverse dimension, quantum mechanically it is fully delocalized and the effective transverse volume in that direction is the whole length of the direction, which is infinite in the non compact case.
The situation is different if the string is classically at a fixed point of a direction, that is it sits at a minimum of a potential.
In this case the quantum delocalization, and so the effective length of the dimension, can be smaller.
The estimate of this effective size, which can be performed explicitly in the study of cosmic strings \cite{jjp}, is a non-trivial task in the present setting.

Finally, the decay rate is computed in flat space. When we go to (weakly) curved spaces, one has to replace $\alpha^\prime$ with an effective $\alpha^\prime_{eff}$ which depends on the warp factors of the metric.\\

Of course, the equation of motion for $\rho(r)$ obtained from (\ref{effact}) is in general not analytically solvable and one must use some numerical or approximated method to evaluate it. We will now consider only the most simple example where we can give an approximate analytical  estimate of the above observable quantities, namely the maximally supersymmetric case $AdS_5\times S^5$ with mesons\footnote{See also \cite{katz2} for a study of meson physics in this setting.} of spin $J\gg \sqrt{\lambda}$ \cite{myers1}, where $\lambda=g_s N$ represents the 't Hooft coupling of the dual theory. In this case the flavor branes are D7-branes and it is convenient to use a different radial coordinate $z=R^2/r$, with $R^4=4\pi\alpha^{\prime 2}\lambda$ such that the relevant part of the metric is given by
\bea
ds^2=\frac{R^2}{z^2}(-dt^2+d\rho^2+\rho^2d\eta^2+dz^2)+\ldots\quad .
\eea
In this case the mass-radius relation is unambiguous and is given by $m_Q=R^2/2\pi\alpha^\prime z_Q$. As discussed in \cite{myers1}, in the case $J\gg\sqrt\lambda$, the spinning string solution is well approximated by a Wilson loop string \cite{malda} slowly spinning around its center of mass, i.e. $\rho(z)\simeq \rho_{st}(z)+\delta\rho(z)$ with very small $\delta\rho(z)$ and
\bea
\rho_{st}(z)=\int_z^{z_0}dx\frac{x^2}{\sqrt{z_0^4-x^4}}\ .
\eea
In this case  $\omega\ll 1$. Also, it is possible to show that in this limit
\bea
\omega^2\simeq  \frac{64{\cal C}^8m_Q^2}{\pi^2\lambda}\left(\frac{\lambda}{J^2}\right)^3 \quad,
\quad z_0^2\simeq \frac{\pi J^4}{16 {\cal C}^6m_Q^2\lambda}\ ,
\eea
 where ${\cal C}=\sqrt2 \pi^{3/2}/\Gamma(1/4)^2\simeq 0.599$.

If we now introduce a lighter  flavor by placing a second D7-brane at a position $z_q< z_0$, the spinning string does intersect it and thus it can split. The condition that the decay is at all possible, which in string theory terms is the fact that the $q$-brane does intersect the string, $z_Q<z_q< z_0$, can be expressed in terms of the particle properties as
\bea\label{mqcritical}
1> \frac{m_q}{m_Q}>\frac{4{\cal C}^3\lambda}{\pi J^2}\ .
\eea
The existence of a limiting minimal value of $m_q$ in order for the decay to happen can be understood in field theory as follows.
Since the theory is Coulombic, the binding energy of the heavy quarks decreases as their distance, and so their spin, increases.
The total energy is $E_{\bar Q Q} \sim 2m_Q -\alpha m_Q\lambda/J^2$, with $\alpha$ being some constant \cite{myers1}.
On the other hand, the binding energy $E^{bind}$ of the meson formed by a heavy and a light quark in the limit of large quark separation is proportional to the mass $m_q$ of the light quark and can be larger than $m_q$ in modulus \cite{katz2,pt}.
The total energy of the two mesons produced in the decay would be $2E_{\bar Q q} \sim
2m_Q - E^{res}$ with positive $E^{res}$.
This is the strong coupling effect that makes it possible for the heavy quark meson to decay, since for large $J^2/m_Q\lambda$ it can be that $E^{res}>\alpha m_Q\lambda/J^2$, so that the total energy of the two produced mesons can be smaller than the one of the heavy quark meson \cite{katz2}.
This is the regime where the string does intersect the $q$-brane in the dual setting.
But, crucially, since $E^{res}= \beta m_q$ for some $\beta$,
for any fixed value of $m_Q\lambda/J^2$, there exists a minimal value of $m_q$ below which $E^{res}$ is smaller than $\alpha m_Q\lambda/J^2$, forbidding the decay.
This critical value
is precisely the one in (\ref{mqcritical}).

As we said before, in order for the decay not to be exponentially suppressed, we must also require that the angular position of the two branes in the transverse direction are equal, $\psi_q=\psi_Q$.
In the dual field theory, an unequal angle $\Delta\psi=\psi_q-\psi_Q\neq 0$ would enter as a phase in the coupling of one type of quark, let us say $q$,  with the complex scalar of ${\cal N}=4$ SYM charged under $\psi$, schematically in the superpotential as ${\cal W}=e^{i\Delta\psi}{\bar q}\Phi q$.
This phase suppresses the decay channel mediated by $\Phi$ and ultimately should be responsible for the exponential suppression of the decay rate.
Of course, a precise and explicit explanation of the suppression in field theory at strong coupling is not possible at present.

Coming back to the string side of the duality, note that the velocity of the string at $z_q$ is of order $\sqrt{\lambda}/J$ and can be neglected in first approximation. Furthermore, if we restrict to the case in which the $q$ D7-brane is not ``too close'' to the $Q$ D7-brane (like for example if $z_Q/z_q=m_q/m_Q\sim \lambda/J^2$), then the decay rate is not completely suppressed since  the angle $\theta$ between the string and the brane is not too close to the value $\pi/2$ and can be evaluated to be
\bea
\theta\simeq{\rm arctg}\sqrt{\Big(\frac{\pi m_q J^2}{4 {\cal C}^3 m_Q\lambda}\Big)^4-1}\ .
\eea
The effective slope is given by $\alpha^\prime_{eff}=\frac{\pi^{-3/2}\sqrt{\lambda}}{2m_q^2}$.

Finally, in order to extract the total decay rate we need the transversal volume $V_\perp$, that in this case is one-dimensional. As we have already said, this is possibly the most subtle point of the whole derivation.
The string is classically at a point of the transverse dimension, so its quantum delocalization can be smaller than the size of the latter.
Contrary to what is done in \cite{jjp}, we cannot estimate the delocalization with a local calculation around the intersection point, since ultimately what generates the classical localization are the boundary conditions on the $Q$-brane, which fix the value of the angle $\psi_Q$ (locally, there is no potential).
So, since the complete calculation of the quantum fluctuations around the classical string embedding is apparently too difficult,
we will adopt a prudent choice that gives as natural (maximal) estimate a transversal length of order  $2\pi R\sim \sqrt{\alpha^\prime}\lambda^{1/4}$.
We expect the actual value of the delocalization to be of the same order.
Then, after taking into account the fact the the string can split at two distinct points, we obtain the following minimal estimate of the decay rate
\bea
\Gamma_{\bar Q Q\rightarrow \bar Q q+\bar q Q}=\frac{m_q\sqrt{\lambda}}{8\sqrt{\pi}N\left(\frac{\pi m_q}{4{\cal C}^3m_Q}\right)^{2}\left(\frac{J^2}{\lambda}\right)^{2}\sqrt{\left(\frac{\pi m_q}{4{\cal C}^3m_Q}\right)^{4}\left(\frac{J^2}{\lambda}\right)^{4}-1}}\ .
\eea

The decay rate has precisely the expected behavior from the field theory point of view. It describes a $1/N$ process that increases as the coupling $\lambda$ increases. As the difference between the mass $m_q$ of the light quark and the mass $m_Q$ of the heavy quark becomes larger and larger, the decay is more and more probable.
However, there is a lower bound on this difference, below which the rate looses its meaning due to the square root. This lower bound is precisely the point at which the $q$ D7-brane reaches the lowest point of the string, below which there is no more intersection and therefore no decay.
Finally, the rate decreases as the spin $J$ of the heavy meson increases.
In fact,
increasing $J$ means increasing the distance between the two heavy quarks and since the theory is non confining, this reduces the binding energy and ultimately the energy density, making the decay process more and more disfavored.

\section{Discussion}\label{sec4}

Let us first discuss the implications of our results in confining models.
In order to be specific, consider the confining theory constructed in \cite{witten,myers2} as a possible dual of large $N$ QCD with a small number of flavors (for which we have Regge-like trajectories \cite{regge3}).
It is a (color) D4 plus (flavor) D6 brane construction, where the branes are wrapped on a supersymmetry breaking circle, leaving us at low energies with 4d large $N$ Yang-Mills and a small number of flavors, coupled to extra KK degrees of freedom.
A first distinctive feature of this model is that there is a lower bound for the effective mass of the quarks.
Even for very small bare mass, the flavor brane cannot reach the horizon.
This is ultimately a consequence of the chiral symmetry breaking.
So, there is a minimum value, different from zero, of the radial position at which the branes can be.
Moreover, for low masses the flavor branes tend to pack as the masses decrease.

\begin{figure}
\begin{center}
\scalebox{0.9}{\includegraphics{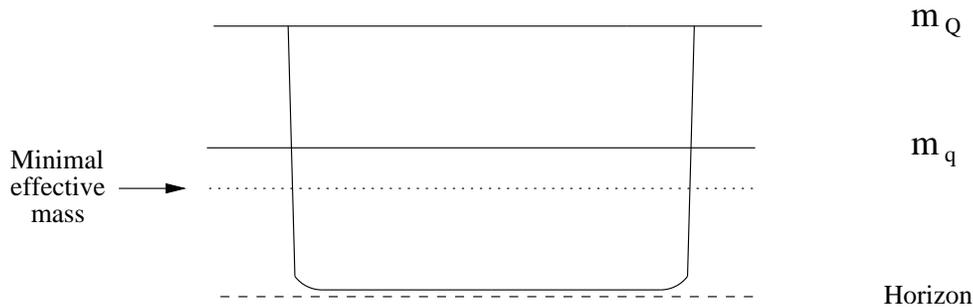}}
\caption{The large spin meson and the light quark brane in the model dual to QCD. The dotted line is the minimal effective mass for the quarks. The string has a ``rectangular'' shape, so the incidence angle on the $q$-brane is approximately $\pi/2$.}
\label{figura2}
\end{center}
\end{figure}

The second salient feature, common to all the confining models, is that in the semi-classical regime of large spin $J$, the string describing a heavy meson is approximately formed by two vertical lines, extending from the brane to the horizon, connected by a horizontal part at (almost) constant (minimal) radius, see figure \ref{figura2}.

If we now introduce a light quark brane, we immediately see that the decay rate is extremely small.
In fact, if on the one hand the new brane, unlike the case of section \ref{sec3}, will always be at a radial position allowing for the intersection with the string, on the other hand it will always intersect it almost perpendicularly.
Since there is a minimum radius for the brane, the latter cannot explore the very small region near the horizon where the string actually turns from vertical to horizontal, where the angle would have been sensibly different from $\pi/2$.
We thus conclude that the effective mass gap for the quarks heavily suppresses the decay of high spin mesons in the strong coupling regime of this model of QCD, and of course in other similar models with the same realization of chiral symmetry breaking \cite{erdmenger}.

Let us now try to give some general interpretation of the string picture of meson decay in field theory terms. Actually, this leads to an
interpretation of the suppression of the splitting as the string and the brane become orthogonal.

The string picture, see for example figure~\ref{figura1}, can be read as a representation of a color flux tube running horizontally
between the positions of the two $Q$-quarks. The string breaks, with the creation of the $q\bar{q}$-pair leading to the decay, at a
specific point on the tube, the position of which depends on the quark masses. Indeed, from the string picture we learn that
this occurs in the vicinity of the $q$-brane, which marks the radial distance corresponding to the energy scale $m_q$.
The corresponding location along the color tube is obtained by projecting
the intersection point on the Minkowski directions.
If we think of the tube as involving a sea of excited gluonic particles,
this suggests that also in the field theory picture this is the point at which
the characteristic energy scale of the gluons in the $\bar QQ$ bound state is of the same order as the light quark masses $m_q$.
The decay rate should increase with  the abundance of such relevant gluonic quanta.

The string picture itself gives us information about this.
The energy density along the color tube, since it involves a projection on the horizontal axis,
is proportional to the slope of the string in figure~\ref{figura1}. This is higher near the quark ends: this corresponds with the fact
that the breaking point with creation of higher mass $q\bar{q}$-pairs is nearer to these ends. At the same time,
the region of the color tube where the
quanta have just the right energy to create a $q\bar{q}$-pair with specific mass becomes smaller as the string is steeper,
eventually vanishing for a position perpendicular to the brane, as for example in figure \ref{figura2}. This may offer a field theory argument why this decay is suppressed.
Note that this almost static reasoning is meant to work only in the large spin limit,
when the spinning string is very extended and spins slowly.
In a more general situation, the dynamics becomes even more nontrivial, and this semiclassical picture seems not to be sufficient.

Another possible decay channel of the mesons is the one involving world-sheet fluctuations that allow the horizontal part of the string to reach the $q$-brane and split.
The study of this kind of process, that is exponentially suppressed with the distance of the $q$-brane to the horizon, that is with the mass $m_q$ of the light quarks, has been initiated in \cite{psz}.
Its importance resides in the fact that it is a stringy analog to the phenomenological picture that underpins the Lund Fragmentation Model for multiparticle production \cite{lund}, which is very successful in describing jet production.
The general quantitative formula we found  may be implemented in this channel too, to calculate the rate of splitting when the string touches the brane, and generalized to other quark processes considered by  the Lund model.
A generalization of the study in \cite{psz} would then allow for an improvement of the string description of the hadronization phase. 
In particular, if confirmed by the data, the intriguing area formula in the phenomenological Lund model begs for a more conceptual explanation from the dual string picture.

Now we comment on the decay of glueballs with high spin. In this case, the  results
concerning the decay of high spin closed strings in flat space, might also be applied to the case of glueballs of confining theories \cite{regge1,regge2}.
The latter are in fact semi-classically described as closed strings sitting at the horizon of the geometry, where they spin in the flat Minkowski part of the background.
The decays we are interested in are the ones that can be described semi-classically, so we concentrate on folded strings, that can decay by splitting at a point.
For circular or elliptical strings this type of decay is exponentially suppressed and in the limit of large mass they are long-lived states \cite{iengo2}.

In the case of the folded strings, the precise matching of the quantum computation \cite{iengo} and the semi-classical one \cite{iengo1} seems to indicate that the decay process in the large mass limit is essentially local and so it should not depend on the details of the transverse space.
If this is really the case, the decay rate of the high spin glueballs described by folded strings could be inferred from the flat space one, with the modifications that the effective string tension $T$ and the dilaton $e^{\Phi}$ at the horizon must be used instead of their flat space counterparts.
In flat space, there are indications that the rate is just a constant and does not depend on the spin of the string \cite{ir}, $\Gamma \sim e^{2\Phi}\sqrt{T}$.
For example, in the Witten model of large $N$ Yang-Mills \cite{witten} the rate in terms of field theory parameters \cite{regge2} would read $\Gamma \sim \frac{m_0\lambda^{7/2}}{N^2}$,
where $\lambda$ is the 't Hooft coupling at the UV cut-off and $m_0$ is the Yang-Mills and Kaluza-Klein mass scale of the theory.
The decay is clearly a $1/N^2$ process that increases with the coupling.

In the case of glueballs one obtains a large phase space for the decay, since the string can split at any point of its length.
The energies and spins of the decay products can be written as functions of this point and semi-classically they are related among each other in a definite way \cite{iengo1}.
The two resulting glueballs are very excited states, not on the leading Regge trajectory.

Finally, one may attempt to estimate other hadron decay rates along the line of this paper, in particular the decay of baryons.
For example, in the ${\cal N}=2$ model of \cite{myers1} the baryons can be seen as $N$ parallel fundamental strings attached to the heavy quark brane and joining a baryon vertex placed at some radial position.
The vertex is a D5-brane wrapped on the five-sphere.
In this picture it is clear that there is no baryon decay if we add a lighter quark brane.
In fact, regardless of its radial position, the brane will be always orthogonal to the fundamental strings, suppressing the decay rate.
In the limit of $N\rightarrow \infty$ a better picture of the baryon may be a spike-like configuration of the heavy quark brane, joining at its tip the baryon vertex.
In this case the light quark brane could intersect the spike at angles different from $\pi/2$, but since the effective description in terms of fundamental strings is no longer valid, the calculation of this paper seems unsuitable to estimate the rate.

\vskip 2cm

\begin{center}
{\large  {\em Acknowledgments}}
\end{center}
We are greatly indebted to F. Bigazzi for his collaboration at some stage of this project.
We would also like to thank M. Bianchi, D. Chialva, R. Iengo, F. Passerini, J. Russo, P. Silva and especially R. Argurio for helpful
discussions, and Cristina Lo Bianco for xfig assistance. This work is supported in part by the FWO - Vlaanderen,
project G.0235.05, by the Federal Office for Scientific, Technical and
Cultural Affairs through the ``Interuniversity Attraction Poles Programme
-- Belgian Science Policy'' P5/27, by the European Community's Human
Potential Programme under contract MRTN-CT-2004-005104 `Constituents,
fundamental forces and symmetries of the universe' and by contracts CYT FPA 2004-04582-C02-01, CIRIT GC 2001SGR-00065.



\end{document}